\documentclass[11pt]{article}
\usepackage{times}

\usepackage{fullpage}
\usepackage{hyperref}       
\usepackage{subcaption}
\usepackage{url}            
\usepackage{booktabs}       
\usepackage{nicefrac}       
\usepackage{microtype}      
\usepackage{amsmath,amsfonts,amssymb,amsthm}
\usepackage{graphicx}
\usepackage{ifthen}
\usepackage{color}

\def\[#1\]{\begin{align}#1\end{align}}
\def\(#1\){\begin{align*}#1\end{align*}}

\newcommand{\bprf}{\begin{proof}}
\newcommand{\eprf}{\end{proof}}
\newcommand{\blem}{\begin{lemma}}
\newcommand{\elem}{\end{lemma}}

\newcommand{\oo}{\mathcal{O}}
\newcommand{\p}[1]{\left(#1\right)}

\DeclareMathOperator{\Var}{Var}

\newcommand{\bP}{\mathbb{P}}
\newcommand{\bI}{\mathbb{I}}
\newcommand{\bE}{\mathbb{E}}

\newcommand{\bF}{\mathbb{F}}

\newtheorem*{theorem*}{Theorem}
\newtheorem{theorem}{Theorem}[section]
\newtheorem{lemma}[theorem]{Lemma}
\newtheorem{conjecture}[theorem]{Conjecture}
\newtheorem{open}[theorem]{Open Problem}
\newtheorem{proposition}[theorem]{Proposition}
\newtheorem{corollary}[theorem]{Corollary}
\theoremstyle{definition}
\newtheorem{definition}[theorem]{Definition}

\DeclareMathOperator{\Ber}{Ber}

\DeclareMathOperator{\Binom}{Binom}

\DeclareMathOperator{\HG}{HG}

\DeclareMathOperator{\PB}{PB}
\newcommand{\rel}{\bowtie}
\newcommand{\DKL}{D_{\mathrm{kl}}}
\newcommand{\DTV}{D_{\mathrm{tv}}}

\usepackage{natbib}
\definecolor{mydarkblue}{rgb}{0,0.08,0.45}                                         
\hypersetup{
  colorlinks=true,
  linkcolor=mydarkblue,
  citecolor=mydarkblue,
  filecolor=mydarkblue,
  urlcolor=mydarkblue,
}
\setcitestyle{authoryear,round,citesep={;},aysep={,},yysep={;}}
\usepackage{tikz}
\usetikzlibrary{patterns,positioning,fit,calc}
\usetikzlibrary{decorations.pathreplacing}
\usepackage{paralist}
\usepackage{enumitem}
\usepackage{mathtools}
\usepackage{algorithm,algorithmicx,algcompatible,algpseudocode}
\usepackage{bbm}

\renewcommand{\paragraph}[1]{\vskip 4pt plus1pt \noindent \textbf{#1}}
\newcommand{\theTitle}{Does robustness imply tractability? A lower bound for planted clique in the semi-random model}
\author{
 \fontsize{11}{13}\selectfont {\bf Jacob Steinhardt}\thanks{\scriptsize Supported by a Fannie \& John Hertz Foundation Fellowship, a NSF Graduate Research Fellowship, and a Future of Life Institute grant.} \\
 \fontsize{11}{13}\selectfont Stanford University \\
 \fontsize{11}{13}\selectfont {\tt jsteinhardt@cs.stanford.edu}
}

\title{\theTitle}

\begin{document}
\algrenewcomment[1]{\hfill $\triangleright$ #1}


\maketitle

\begin{abstract}
We consider a robust analog of the planted clique problem. In this analog, a set 
$S$ of vertices is chosen and all edges in $S$ are included; then, edges between 
$S$ and the rest of the graph are included with probability $\frac{1}{2}$, while 
edges not touching $S$ are allowed to vary arbitrarily.
For this \emph{semi-random} model, 
we show that the information-theoretic threshold for recovery is $\tilde{\Theta}(\sqrt{n})$, 
in sharp contrast to the classical information-theoretic threshold of $\Theta(\log(n))$. 
This matches the conjectured computational threshold for the classical planted clique 
problem, and thus raises the intriguing possibility that, once we require robustness, 
there is no computational-statistical gap for planted clique. 
Our lower bound involves establishing a result regarding the KL divergence of a family of 
perturbed Bernoulli distributions, which may be of independent interest.
\end{abstract}

\thispagestyle{empty}
\newpage
\setcounter{page}{1}

\section{Introduction}

The planted clique problem is perhaps the most famous 
example of a \emph{computational-statistical gap}---while 
it is information-theoretically possible to recover planted cliques 
even of size $2\log_2(n)$, the best efficient recovery algorithms 
require the clique to have size $\Omega(\sqrt{n})$. 
It has long been conjectured that no polynomial-time algorithm can 
find cliques of size $n^{1/2-\epsilon}$, with recent breakthrough 
work by \citet{barak2016nearly} establishing this for the class of 
sum-of-squares algorithms. There thus appears to be an exponential gap between 
what is possible statistically and what is tractable computationally.


In this paper we revisit this gap, and question whether recovering cliques of 
size $s \ll \sqrt{n}$ is actually meaningful, or if it can only be done by 
over-exploiting the particular details of the planted clique model.
Recall that in the planted clique model, a set $S$ of vertices is 
chosen at random and all vertices in $S$ are connected; the remaining edges are then each included 
independently with probability $\frac{1}{2}$. 
While this model is convenient in its simplicity, it also has a number of 
peculiarities---for instance, simply returning the highest-degree 
nodes already performs nearly as well at recovering $S$ as sophisticated spectral algorithms.

\citet{feige2001heuristics} argue that it is more realistic to consider a 
\emph{semi-random} model, in which edges that 
do not touch any vertices in the clique are allowed to vary arbitrarily. 
This forces recovery algorithms to be more robust by not relying 
on simple heuristics such as maximum degree to identify the planted clique.
It is then natural to ask---once we require such robustness, how large 
must a clique be to be statistically identifiable?
To this end, we establish a strong information-theoretic lower bound:

\begin{theorem*}
In the semi-random model, it is information-theoretically impossible to 
even approximately recover 
planted cliques of size $o(\sqrt{n})$. Moreover, it is information-theoretically 
possible to exactly recover cliques of size $\omega(\sqrt{n\log(n)})$.
\end{theorem*}

It is interesting that the information-theoretic threshold in the semi-random model 
essentially matches the computational threshold of $\sqrt{n}$ in the standard model. 
It is tempting to hope that, in the semi-random model, the computational and statistical 
thresholds in fact coincide---i.e., that there is an efficient algorithm for recovering 
cliques of size $\sqrt{n\log(n)}$. 
In such a case, the previous exponential gap between the statistical and computational limits 
would vanish entirely. 

The best upper bound we are aware of is via the results of 
\citet{charikar2017learning} on robust estimation in the $\ell_2$-norm, which allows 
recovery of cliques of size $\tilde{\Omega}(n^{2/3})$. While this does not match 
the $\tilde{\Theta}(\sqrt{n})$ information-theoretic threshold, in a sense this is a much
smaller gap than before---in the non-robust case there was an exponentially large gap 
of $\log(n)$ versus $\sqrt{n}$, whereas here the gap is now at most $\sqrt{n}$ versus 
$n^{2/3}$. Moreover, the algorithm in \citet{charikar2017learning} solves a much more 
general problem and is not tailored to the planted clique problem, so there is reason to hope 
that the true gap is smaller.
Since initial publication of this paper, \citet{mckenzie2018robust} improved this upper bound 
to $\oo(n^{2/3})$, removing the log factor.

\subsection*{The Model}

We now explain the semi-random model in more detail. 
%
%
We consider a graph on $n$ nodes with a planted clique of size $s$. 
Label the nodes $1, \ldots, n$, and let $S \subseteq \{1,\ldots,n\}$ be the set of vertices 
in the clique. We represent the graph by its adjacency matrix $A \in \{0,1\}^{n \times n}$, 
which must be symmetric and satisfy $A_{ii} = 0$ for all $i$. Beyond these constraints, 
$A$ is generated as follows:
\begin{equation}
A_{ij} = \left\{ \begin{array}{ccl} 1 & : & i, j \in S \\ \Ber(\frac{1}{2}) & : & i \in S, j \not\in S \text{ or } j \in S, i \not\in S \\ \text{arbitrary} & : & \text{else} \end{array} \right.
\end{equation}
Here $\Ber(p)$ denotes the Bernoulli distribution with parameter $p$ (i.e., a coin toss 
with probability $p$ of being $1$). In words, $A$ is generated by planting a clique 
$S$ and connecting all pairs of vertices in $S$; then connecting each clique vertex to each 
non-clique vertex independently with probability $\frac{1}{2}$; and finally filling in edges 
between non-clique vertices arbitrarily. 

This is essentially the model considered by \citet{feige2001heuristics}, except they 
additionally allow any number of edges between $S$ and $[n]\backslash S$ to be deleted.
The focus of this paper is on lower bounds, so restricting the edges between $S$ and 
$[n] \backslash S$ to be random (i.e., disallowing deletions among these edges) 
as above will only make proving lower bounds 
harder---in particular, any lower bounds in our model automatically hold in 
\citeauthor{feige2001heuristics}'s model.
Our information-theoretic upper bound also holds in 
\citeauthor{feige2001heuristics}'s model, though we will give an algorithmic upper 
bound that does require the edges between $S$ and $[n] \backslash S$ to be random.
For the most part, however, a reader who prefers to think about 
\citeauthor{feige2001heuristics}'s model can do so.

\subsection*{Motivation, Background, and Related Work}

While the planted clique model is a central tool for understanding 
average-case hardness, it has long been understood that it is somewhat artificial as a 
model of real data, as it imposes strong regularities on the graph asserting 
that every vertex has a neighborhood that is either completely random or completely random 
apart from the clique. This fails to capture interesting structure in real graphs---for instance, 
we might think of planted clique as a stylized model for community detection, 
where the goal is to statistically recover a community of highly connected 
vertices in a social network. In such a case, we might expect to see high-degree ``hub'' vertices 
(which, despite their high degree, may not be part of a community) 
as well as heterogeneity among different vertices. Both phenomena are ruled out by the 
classical planted clique model.


To deal with this, \citet{blum1995coloring} proposed the 
\emph{monotone adversaries}\footnote{\citeauthor{blum1995coloring}'s model has been alternately 
referred to as the monotone adversaries model or the semi-random model, while 
\citeauthor{feige2001heuristics}'s model has only been referred to as the semi-random model. 
We will always refer to the former as the monotone adversaries model and the latter as the 
semi-random model in this paper.} model; in this model, a graph is first drawn from the classical 
planted clique model, and then an adversary is allowed to remove any of the non-clique edges. 
Intuitively, this should only make the problem easier; however, this model already 
re-assuringly rules out 
simple degenerate heuristics (such as returning the highest-degree nodes) that work in the 
original planted clique model but not the monotone adversaries model.

While the monotone adversaries model does allow interesting variation beyond the 
original planted clique model, it is still quite restrictive---it only allows the creation 
of graphs with a single large clique, while real-world graphs typically have many 
densely connected communities.
One could explicitly extend the model to have multiple planted cliques (or more generally 
multiple planted dense subgraphs, as in \citet{chen2014improved}, \citet{guedon2014community}, 
and \citet{agarwal2015multisection}), but this is still 
unsatisfying---it forces us to model the entire graph just to learn about a single 
clique/community. If our model for the rest of the graph is wrong, then formally, any 
recovery guarantees will no longer hold.

This leads us to \citeauthor{feige2001heuristics}'s model as described above, 
which allows not only 
edge deletions, but also edge \emph{additions} as long as the added edges have both 
endpoints outside the planted clique.
Rather than having to explicitly model additional cliques, the additional cliques 
are elegantly accounted for as edges that may be added in the rest of the graph. 
Beyond modeling a richer family of graph structures, 
\citeauthor{feige2001heuristics}'s model is also interesting because it gets at the 
notion of \emph{which structures can be recovered locally} (i.e., by relying only on 
the planted structure and its neighborhood).

While this model seems very natural to us, it unfortunately does not appear very 
well-studied, despite a great deal of work on the weaker monotone adversaries model 
\citep{feige2000finding,coja2004coloring,
chen2014improved,guedon2014community,moitra2015robust,agarwal2015multisection}. 
In fact, to our knowledge, only two papers since \citeauthor{feige2001heuristics} have 
studied versions of their model---\citet{coja2007solving}, who study maximum independent 
sets and $k$-colorings, and \citet{makarychev2012approximation}, who propose a 
fairly general model for semi-random graph partitioning.
The paucity of results in this model is perhaps because, while semidefinite programs 
form a natural class of algorithms 
that work in the monotone adversaries model, it appears challenging to design algorithms 
for the full semi-random model. 
Moreover, from a statistical perspective, it might intuitively 
seem impossible to recover the planted clique in the face of such a powerful source of 
adversarial corruptions.

Some recent work on learning in the presence of adversarial corruptions has caused us 
to question this intuition, and to feel that the semi-random model deserves renewed interest. 
Specifically, \citet{charikar2017learning} study learning in the presence of a large 
(possibly even majority) fraction of corruptions. 
Among other settings, they propose a model of clustering where, rather than thinking of 
the data as a mixture of $k$ clusters, one only models a single cluster and allows the data 
that would have come from the remaining $k-1$ clusters to instead be completely arbitrary. 
Surprisingly, it is possible to efficiently and robustly cluster the data essentially as well 
as is possible with the best non-robust computationally efficient algorithms. While one 
might have initially suspected that such a strong adversary would destroy any hope of 
statistical recovery, in fact we lose nothing as long as we restrict attention to 
efficient algorithms.

\citet{charikar2017learning} also study a robust version of the stochastic block model, 
which is a generalization of the model studied here. In the present setting, 
their results imply that it is possible to robustly recover planted cliques of size 
$\tilde{\Theta}(n^{2/3})$ in polynomial time; this puts to rest the idea that 
efficient algorithms are unattainable in this setting. This, together with the 
observation (proved below as Theorem~\ref{thm:main-upper}) that it is 
\emph{information-theoretically} possible to recover cliques of size 
$\tilde{\Theta}(\sqrt{n})$--i.e., at the same threshold as algorithmic 
recoverability in the classical model--led us to consider the 
semi-random model much more closely. 

Here, we obtain a better understanding of this model by showing that the information-theoretic 
recovery threshold is \emph{exactly} $\tilde{\Theta}(\sqrt{n})$. 
This exhibits a large gap relative to the monotone adversaries model (for which recovery is 
possible at size $\Theta(\log n)$); it also establishes the threshold at which statistical 
recovery of the clique is possible from purely local information (i.e., from the clique and its 
neighborhood). 
While recovery below this threshold may 
be possible, it would necessarily involve assumptions on parts of the 
graph that are disjoint from the planted clique.

\paragraph{Other related work.} 
\citet{moitra2015robust} recently studied information-theoretic thresholds 
for planted recovery in the monotone adversaries model, and established the surprising result 
that even for monotone adversaries the \emph{two-community stochastic block model} 
problem is information-theoretically harder than in the absence of adversaries. 
The gap shown there is only a constant factor, in contrast to the $\log n$ vs. 
$\sqrt{n}$ gap shown here (though of course the models are different and so the results 
are incomparable).

\citet{steinhardt2018resilience} recently established 
information-theoretic upper bounds for a large class of robust learning problems, 
and our upper bound here follows their strategy. 
Non-trivial information-theoretic \emph{lower bounds} for robust 
recovery are still somewhat lacking, and the results here provide one step towards that end.

\subsection*{Recovery Criterion} 

In the semi-random model, it is necessary to modify our notion of what it means to recover 
the planted clique $S$: 
Because edges not touching $S$ can be chosen arbitrarily, the set $S$ need not be the unique 
large clique in the graph, even when the clique size $s$ is very large. 
Indeed, even if $s = \frac{n}{2}$, we could 
observe a graph with two cliques of size $\frac{n}{2}$ (with all edges between them occurring 
with probability $\frac{1}{2}$) and it would be impossible to tell which of the two was the 
planted clique $S$.

How can we overcome this ambiguity? One natural condition is to ask that 
\emph{most members of $S$ are able to identify $S$}. 
In other words, if we think of the clique as a community, members of that community should be 
able to identify the other members.

We can formalize this as follows: we are given the adjacency matrix $A$ of the graph 
and a vertex $v$ sampled uniformly from $S$, and asked to output an estimate $\hat{S}(A,v)$ 
of the set $S$. 
(The vertex $v$ is sampled after $A$ is generated, i.e. 
the adversary in the semi-random model is not allowed to respond to the choice of $v$.)
This is a special case of the \emph{semi-verified} model introduced by 
\citet{charikar2017learning}. 
We succeed if $S$ and $\hat{S}$ are similar according to some similarity measure.


Our specific similarity measure is the \emph{Jaccard index} 
$J(S,\hat{S}) = \frac{|S \cap \hat{S}|}{|S \cup \hat{S}|}$, although our results are 
mostly agnostic to this choice. 
We will show that when the clique size $s \ll \sqrt{n}$, even partial recovery is 
impossible under this 
measure---$J(S,\hat{S}) = o(1)$ for all estimators $\hat{S}$ whenever $s = o(\sqrt{n})$.
In fact, we will show that it is impossible to distinguish between the case where a 
subset of $s$ vertices belongs to a unique planted clique, and the case where every vertex 
belongs to $\Theta(\sqrt{n}/s)$ planted cliques. In contrast, when $s \gg \sqrt{n\log(n)}$, 
\emph{exact} recovery is possible---$J(S,\hat{S}) = 1$ with high probability.

We note that some other work on the semi-random model, such as 
\citet{makarychev2012approximation}, considers a different notion of success. Rather than 
asking to recover the specific clique $S$, they are happy with recovering \emph{any} large 
clique. This is a reasonable notion if one's goal is to \emph{test the effectiveness of 
graph heuristics}, i.e. to use the semi-random model as a rich test-bed for analyzing 
heuristics for maximum clique and other NP-hard problems. That goal is certainly both 
well-motivated and interesting, but is not the focus here. From a statistical 
perspective, identifying the specific clique $S$ is in fact important---if $S$ is 
unidentifiable, then the original statistical recovery problem was ill-posed to begin with, 
and no heuristic can circumvent this.

\subsection*{Results and Techniques}

Our main result is an information-theoretic lower bound on recovery in the semi-random model:
%
\begin{theorem}
\label{thm:main-lower}
For any $n$ and $s$, there is a distribution over instances of the semi-random 
planted clique model, each with planted clique size at least $s$, 
such that given the adjacency matrix $A$ and $v$ sampled uniformly from $S$, any candidate 
$\hat{S}(A, v)$ for $S$ must satisfy
\begin{equation}
\bE_{A,v}\left[J(S,\hat{S}(A,v))\right] \leq \oo\Big(\frac{s+\log(n)}{\sqrt{n}}\Big).
\end{equation}
\end{theorem}
In particular, this means that the Jaccard similarity goes to $0$ whenever $s/\sqrt{n}$ goes to $0$.

The proof idea is to construct a ``null distribution'' $P_0(A)$ under 
which every vertex 
belongs to exactly $k$ identically distributed cliques,
and show that it has small total variational distance to an instance $P_1$
of the semi-random model. Then, given $v$, it is impossible to tell which of the $k$ cliques 
is $S$, and so one cannot achieve Jaccard similarity better than $\oo(\frac{1}{k})$.
It turns out we can take $k = \Theta(\sqrt{n}/s)$ while still maintaining low total 
variational distance.

\paragraph{Lower bound construction.} 
One challenge is how to construct the $k$ identically distributed cliques. Ideally, we would 
like every pair of cliques to have small intersection, since in the semi-random model the 
cliques are supposed to be disjoint. 
We take inspiration from a well-known combinatorial design involving finite fields. 
Specifically, for any finite field $\bF_m$, we can consider the line 
in $\bF_m^2$ with a given slope $r$ and intercept $h$, i.e. 
$L_{r,h} = \{(a,b) \mid a-br = h\}$. If we take $L_{r,h}$ where $r$ ranges from 
$0$ to $k-1$ and $h$ ranges from $0$ to $m-1$, then every point in $\bF_m^2$ is covered by 
exactly $k$ lines, and every pair of lines intersects in at most $1$ point. We will sample 
points from $\bF_m^2$ without replacement, and create a clique for each of the 
lines $L_{r,h}$.

\paragraph{Coupling and perturbed Bernoulli distributions.}
Using a coupling argument, we will show that the above construction $P_0$ can be approximately 
simulated using an instance $P_1$ of the semi-random model. However, a difficulty is that 
$P_0$ has a number of dependencies (due to the additional cliques) that cannot exist in 
$P_1$ due to the independence between 
the random edges in $P_1$. To address this, we establish a technical 
result--Proposition~\ref{prop:kl-multi-2}--regarding a family of 
``perturbed Bernoulli'' distributions, 
which shows that their KL divergence is small as long as the ``low-order statistics'' of the 
perturbations are similar. In our context, this shows that the dependencies in $P_0$ created 
by the additional cliques are sufficiently hidden by the randomness in the remaining edges.
The family of perturbed Bernoulli distributions is quite general and this bound 
may be useful in other cases, as well.

\paragraph{Upper bound.} 
We also establish an upper bound 
showing that exact recovery is 
information-theoretically possible for sets of size $\sqrt{n\log(n)}$, and computationally 
possible for sets of size $n^{2/3}\log^{1/3}(n)$:

\begin{theorem}
\label{thm:main-upper}
Under the semi-random model, for a planted clique 
of size $s = \omega(\sqrt{n\log(n)})$, exact recovery is possible with high probability---
there is an $\hat{S}(A,v)$ such that $\bP_{A,v}[S = \hat{S}(A,v)] = 1 - o(1)$.

Moreover, there exists a constant $C$ such that for $s \geq C n^{2/3}\log^{1/3}(n)$, 
\emph{efficient} exact recovery is possible---there is a polynomial-time algorithm $\hat{S}(A,v)$ 
such that $\bP_{A,v}[S = \hat{S}(A,v)] = 1 - o(1)$.
\end{theorem}
This result is relatively straightforward but we include it for completeness.
The information-theoretic bound uses the fact that the true planted clique must have 
small intersection (size at most $O(\log n )$) 
with any other large clique, and therefore for cliques of size $s$ we can recover 
all but a $\frac{n\log(n)}{s^2}$ fraction of the vertices in the clique 
(this follows an argument of \citet{steinhardt2018resilience}). 
We note this bound holds in \citeauthor{feige2001heuristics}'s model as well (i.e., it 
still holds under deletions of edges between $S$ and $[n] \backslash S$).
The computational bound is essentially Corollary 9.3 of \citet{charikar2017learning}; unlike 
the information-theoretic bound, it is not robust to deletions between $S$ and $[n] \backslash S$.

\subsection*{Open Problems}

The results in this paper show that in the semi-random model, the information-theoretic 
threshold for planted clique is between $\sqrt{n}$ and $\sqrt{n\log(n)}$.
We believe that the correct threshold is $\sqrt{n\log(n)}$:

\begin{conjecture}
\label{conj:lower}
If $s = o(\sqrt{n\log(n)})$, no algorithm can even partially recover $S$ in the semi-random model. 
\end{conjecture}
In our lower bound construction below, we consider distributions over cliques that overlap in a 
single element. Proving Conjecture~\ref{conj:lower} would require considering cliques 
that overlap in $\log(n)$ elements, which seems possible but more challenging (note that we show 
in the proof of Theorem~\ref{thm:main-upper} that it is impossible for cliques to overlap in more 
than $\oo(\log n)$ elements).

Computationally, our upper bound is at 
$s = \tilde{\Theta}(n^{2/3})$, while in the purely random model there are algorithms 
which succeed at $s = \Theta(\sqrt{n})$. This is a large gap, and one might hope to do better, 
especially given that the computational upper bound is tailored to a much more general problem 
that does not exploit the planted clique structure.
\begin{open}
Is there a polynomial-time algorithm that, for $s = \omega(\sqrt{n\log(n)})$, 
exactly recovers $S$ with high probability under the semi-random model?
\end{open}
A resolution in either direction would be interesting---if resolved in the positive, 
this would mean that, once we require robustness, the computational-statistical gap for 
planted clique vanishes; on the other hand, if resolved in the negative, it would yield 
an instance where there is a computational gap between robust recovery and non-robust recovery.


Finally, we can generalize the planted clique model to the more general 
\emph{planted dense subgraph} model, in which edges within $S$ are formed with probability 
$p$, and edges between $S$ and $[n] \backslash S$ are formed with probability $q$. The 
planted clique model corresponds to $p = 1$, $q = \frac{1}{2}$. We can then ask for the 
information-theoretic recovery threshold in this more general model, as a function of 
$p$ and $q$. 

\begin{conjecture}
\label{conj:dense-subgraph}
The information-theoretic threshold for recovery in the semi-random planted dense subgraph 
model is $\tilde{\Theta}(\sqrt{\frac{n(p+q)}{(p-q)^2}})$.
\end{conjecture}

The threshold in Conjecture~\ref{conj:dense-subgraph} is the 
\emph{Kesten-Stigum threshold} \citep{kesten1966limit,kesten1966additional}, 
which is believed to be the threshold for
efficient recoverability in the stochastic block model 
\citep{decelle2011asymptotic}. 
As semi-random dense subgraph is a natural robust analog of the stochastic block 
model, a proof of Conjecture~\ref{conj:dense-subgraph} would be another data point 
suggesting a relationship 
between robustness and computation. We remark that at least the upper bound 
can be established with similar ideas
to Theorem~\ref{thm:main-upper}. 

\section{Perturbed Bernoulli Distributions}
\label{sec:pb}

The main technical tool in our proof is a bound on a 
class of distributions that we refer to as \emph{perturbed Bernoulli} distributions. 
These are obtained by first sampling $s$ variables $X_{1:s} \in \{0,1\}$ independently 
from a Bernoulli distribution, then setting some subset of them to $1$. More formally, 
consider the following process:
\begin{itemize}
\item First, $Y_1, \ldots, Y_s \sim \Ber(q)$.
\item Then, a set $J \subseteq \{1,\ldots,s\}$ is sampled according to a probability mass 
      function $\sigma$ (i.e., $\sigma(J)$ is the probability of sampling $J$).
\item Finally, $X_j$ is set to $1$ if either $Y_j = 1$ or $j \in J$.
\end{itemize}

\begin{definition}
\label{def:pb}
We say that a random variable distributed as $X_{1:s}$ above 
is a \emph{perturbed Bernoulli distribution} with parameters 
$q$ and $\sigma$. We denote it by $X_{1:s} \sim \PB(q, \sigma)$.
\end{definition}

In our analysis in the subsequent sections, we will take 
$q \approx \frac{1}{2}$, and $\sigma$ will place most of 
its mass on sets of size $0$ or $1$. 
It turns out that many of the distributions we will want to 
analyze in comparing the semi-random model to a null model 
are perturbed Bernoulli distributions, where intuitively we would 
like to say that a perturbed Bernoulli distribution is close to 
the original $\Ber(q)$ distribution. Unfortunately, such a comparison 
is too coarse, and we instead need to show that two perturbed Bernoulli 
distributions are close if the perturbations have ``similar low-order statistics''. 
This can be made precise in the following proposition:

\begin{proposition}
\label{prop:kl-multi-2}
\label{prop:pb}
Suppose that $X_{1:s} \sim \PB(q, \sigma)$, and $X_{1:s}' \sim \PB(q, \tau)$. 
Let $S(J) = \sum_{J' \supseteq J} \sigma({J'})$ and define $T(J)$ similarly 
in terms of $\tau$. Then, we have 
\begin{equation}
\label{eq:kl-multi-2}
\DKL(P_{X}, P_{X'}) \leq \frac{1}{\tau({\emptyset})} \sum_{J \subseteq \{1,\ldots,s\}} \Big(\frac{1-q}{q}\Big)^{2|J|} (S(J) - T(J))^2,
\end{equation}
where $P_X$ and $P_{X'}$ are the distributions over $X$ and $X'$ and $\DKL$ denotes KL divergence.
\end{proposition}
To interpret Proposition~\ref{prop:kl-multi-2}, first note that $\tau({\emptyset})$, 
the probability that $J$ is the empty set under $\tau$, must be non-negligible 
for \eqref{eq:kl-multi-2} to be meaningful. We will specifically have in mind the 
case where $\sigma$ and $\tau$ place most of their mass on sets of small size (say, for 
simplicity, sets of size at most $1$). To use Proposition~\ref{prop:kl-multi-2} 
in this case, one would first argue that $\sum_{|J| \geq 2} (\frac{1-q}{q})^{2|J|} (S(J) - T(J))^2 \leq 2\sum_{|J| \geq 2} (\frac{1-q}{q})^{2|J|} (S(J)^2 + T(J)^2)$ 
is negligible, and next argue that $S(J)$ and $T(J)$ take similar values on sets of 
size $1$ (the size $0$ case is trivial because $S({\emptyset}) = T({\emptyset}) = 1$).
This is useful because it allows us to show that two perturbed Bernoulli distributions 
are similar as long as the ``low-order statistics'' of $\sigma$ and $\tau$ match and the 
``high-order statistics'' are negligible.

The proof of Proposition~\ref{prop:kl-multi-2} is given in Appendix~\ref{sec:kl-multi-2-proof}.
It is based on upper-bounding KL divergence by $\chi^2$ divergence, and then analyzing a 
sort of Fourier transform of the probability distributions $P_X$ and $P_{X'}$.

As a final note, if $q' > q$ then sampling from a $\Ber(q')$ distribution is equivalent 
to sampling from $\PB(q, \sigma)$ with $\sigma(J) = (\frac{q'-q}{1-q})^{|J|}(\frac{1-q'}{1-q})^{s-|J|}$. This is because flipping a coin with weight $q'$ is equivalent to 
flipping coins with weights $q$ and $\frac{q'-q}{1-q}$, and calling the result heads if 
if either coin comes up heads.

\section{Lower Bound: Warm-Up}
\label{sec:lower}

In this section we will prove a weaker version of Theorem~\ref{thm:main-lower}, 
showing that the probability of recovering $S$ cannot be much larger than $\frac{1}{2}$ 
(the proof of the full version of Theorem~\ref{thm:main-lower} 
is given in Section~\ref{sec:lower-full}). 
The idea is to construct a null distribution $P_0$ which does not lie 
in the semi-random model, and from which $S$ cannot be recovered, 
and show that it is close to a distribution $P_1$ lying in the semi-random model.

\paragraph{The null distribution $P_0$.}
We start by describing $P_0$, which is parameterized 
by integers $n$ (the size of the graph) and $m$; 
it will induce cliques of size roughly $\frac{n}{m}$.
It is constructed so that samples from $P_0$ have the following properties:
\begin{itemize}
\item The graph contains $2m$ cliques, each of size approximately $\frac{n}{m}$.
\item Every vertex lies in two cliques.
\item Every pair of cliques is either disjoint or has intersection $1$.
\end{itemize}
We will do this as follows, creating two groups of $m$ cliques each.
For each vertex $i \in [n]$, sample (without replacement) a pair $(a_i, b_i) \in [m] \times [m]$. 
Vertex $i$ will belong to the $a_i$th clique in the first group, and the $b_i$th clique 
in the second group. Note that sampling without replacement ensures that no two cliques 
intersect in more than one element. 

Given $a_{1:n}$ and $b_{1:n}$, we generate the adjacency matrix $A$ for the graph as follows:
\[ A_{ij} &= \left\{ \begin{array}{ccl} 1 & : & a_i = a_j \text{ or } b_i = b_j, \\ \Ber(q) & : & \text{ else.} \end{array} \right. \]
Here $q$ will be chosen to be $\frac{1}{2} - \frac{1}{2m-2}$; 
it is slightly less than $\frac{1}{2}$ in order to correct for the extra edges created by 
the two cliques, so that the expected degree is the same as in the semi-random model.

\paragraph{The semi-random instance $P_1$.}
We want to construct a semi-random instance $P_1$ that is close to $P_0$. We
will do this via a coupling argument: we express a sample from $P_0$ as a sequence of 
local decisions, and show that each decision can either be exactly imitated under $P_1$, or 
approximately imitated with small KL divergence.

First let us re-express $P_0$ in a way that makes it look more like a planted clique instance, where 
we will think of the clique corresponding to $a_i = 1$ as the ``planted'' clique $S$.
Let $\HG(n,K,N)$ denote the hypergeometric distribution, which samples $n$ items 
without replacement from $[N]$, and counts the number of sampled items lying in $\{1,\ldots,K\}$.
We can generate $A \sim P_0$ sequentially as follows:
\begin{enumerate}
\item Sample $s \sim \HG(n,m,m^2)$ and let $S$ be a uniformly random subset of $[n]$ of size $s$. 
      Set $a_i = 1$ for each $i \in S$ and sample the corresponding $b_i$ from $[m]$ without replacement.
      Connect all vertices in $S$.
\item For each $i \not\in S$, determine which edges should exist between $S$ and 
      $[n] \backslash S$. This involves sampling $(a_i,b_i)$ and determining if 
      $b_i = b_j$ for some 
      $j \in S$ (in which case $A_{ji} = 1$), and including the remaining edges 
      with probability $q$.
\item Finally, fill in all of the remaining edges (i.e., the edges between 
      elements of $[n] \backslash S$) conditioned on the decisions made in steps $1$ and $2$.
\end{enumerate}

Steps $1$ and $3$ can be exactly mimicked under the semi-random model---step $1$ 
because it involves planting a clique at random, and step $3$ because it only involves 
decisions in $[n] \backslash S$, which we are allowed to choose arbitrarily.
For step $2$, however, we cannot exactly mimic $P_0$ in the semi-random model---our 
hands are tied because all edges between $S$ and $[n] \backslash S$ must be generated at random. 
%
%
Under $P_1$, then, steps $1$ and $3$ are the same as $P_0$, but we use the alternate step $2'$:
\begin{enumerate}
\item[$2'$.] For each $i \not\in S$, determine which edges should exist between $S$ and $[n] \backslash S$. 
             This involves setting $A_{ji}$ to be $1$ with probability $\frac{1}{2}$ independently for all $j \in S$.
\end{enumerate}
It is also necessary to sample $(a_i, b_i)$ in step $2'$, since these are used in step $3$. 
Since $a_i$ and $b_i$ do not affect the edges generated in $2'$, we can sample them however we 
want without violating the semi-random model. 
For each $i$, we will thus mimic $P_0$ by sampling $(a_i,b_i)$ from the conditional distribution under $P_0$ given 
the decisions made so far (e.g. conditioned on $S$, on the previous $(a_{i'}, b_{i'})$, and on $A_{ji}$). 

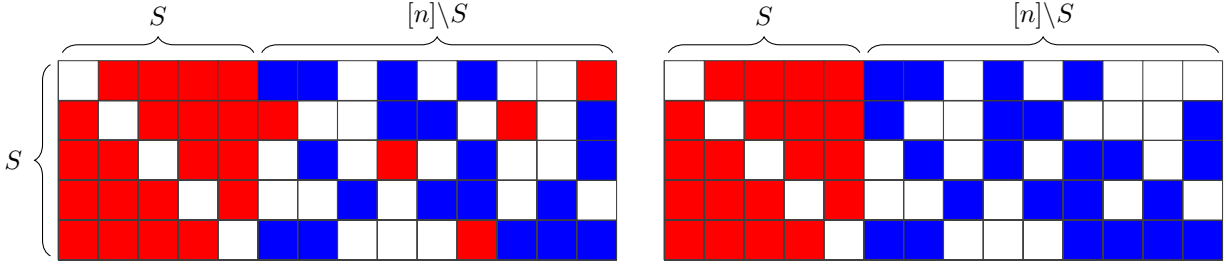
\begin{figure}
\def\pixelmap{{%
{2,2,2,2,0,1,1,0,0,0,2,1,1,1},
{2,2,2,0,2,0,0,1,0,1,1,0,1,0},
{2,2,0,2,2,0,1,0,2,0,1,0,0,1},
{2,0,2,2,2,2,0,0,1,1,0,2,0,1},
{0,2,2,2,2,1,1,0,1,0,1,0,0,2}
}}

\def\pixelmapii{{%
{2,2,2,2,0,1,1,0,0,0,1,1,1,1},
{2,2,2,0,2,0,0,1,0,1,1,0,1,0},
{2,2,0,2,2,0,1,0,1,0,1,1,0,1},
{2,0,2,2,2,1,0,0,1,1,0,0,0,1},
{0,2,2,2,2,1,1,0,1,0,1,0,0,0}
}}

\colorlet{color0}{white}
\colorlet{color1}{blue}
\colorlet{color2}{red}
\colorlet{color0.0}{white}
\colorlet{color1.0}{blue}
\colorlet{color2.0}{red}

\begin{center}
\begin{tikzpicture}[grid/.style={thin,gray!50!black}, scale=0.53]

\foreach \y [count=\i from 0] in {0,1,...,4}
    \foreach \x [count=\j from 0] in {0,1,...,13}
    {
        \def\bit{0}
            \pgfmathsetmacro{\bit}{\pixelmap[\i][\j]}
        \fill[color\bit] (\x,\y) -- +(0, 1) -- +(1, 1) -- +(1,0) -- cycle;
        \draw[grid] (\x,0) -- (\x,5);
        \draw[grid] (0,\y) -- (14,\y);

        \def\bitii{0}
            \pgfmathsetmacro{\bitii}{\pixelmapii[\i][\j]}
        \fill[color\bitii] (\x+15.2,\y) -- +(0, 1) -- +(1, 1) -- +(1,0) -- cycle;
        \draw[grid] (\x+15.2,0) -- (\x+15.2,5);
        \draw[grid] (0+15.2,\y) -- (14+15.2,\y);
    }
\draw[grid] (14,0) -- (14,5);
\draw[grid] (0,5) -- (14,5);
\draw [decorate,decoration={brace,amplitude=6pt}] (0.1+15.2,5.2) -- (4.9+15.2,5.2) node[midway,yshift=14pt] {\small $S$};
\draw [decorate,decoration={brace,amplitude=6pt}] (5.1+15.2,5.2) -- (13.9+15.2,5.2) node[midway,yshift=14pt] {\small $[n] \backslash S$};
\draw [decorate,decoration={brace,amplitude=6pt}] (0.1,5.2) -- (4.9,5.2) node[midway,yshift=14pt] {\small $S$};
\draw [decorate,decoration={brace,amplitude=6pt}] (5.1,5.2) -- (13.9,5.2) node[midway,yshift=14pt] {\small $[n] \backslash S$};
\draw [decorate,decoration={brace,amplitude=6pt}] (-0.2,0.1) -- (-0.2,4.9) node[midway,xshift=-14pt] {\small $S$};
\draw[grid] (14+15.2,0) -- (14+15.2,5);
\draw[grid] (0+15.2,5) -- (14+15.2,5);
\end{tikzpicture}
\end{center}
\caption{Illustration of step $2$ of $P_0$ (left) vs. step $2'$ of $P_1$ (right). Each 
grid is the top $s \times n$ portion of the adjacency matrix. Red edges are from 
cliques while blue edges are random. In the left matrix, $b_6 = 2$, $b_9 =  3$, $b_{11} = 5$, $b_{12} = 2$, and $b_{14} = 1$ (the 
remaining elements of $b_{6:14}$ are all larger than $5$). The extra red edges introduce anti-correlations in $P_0$ 
which could potentially distinguish it from $P_1$. We will bound these 
correlations using Proposition~\ref{prop:pb}.}
\label{fig:step2}
\end{figure}

\paragraph{Comparing $P_0$ and $P_1$.}
We next want to show that $P_0$ and $P_1$ are close. For this, it helps to more concretely 
represent the difference between $P_0$ and $P_1$ in step $2$. For convenience, number 
the vertices in $S$ as $1,\ldots,s$ and assume that $b_i = i$ for $i = 1,\ldots,s$. 
Suppose that we have already filled in the edges from $S$ to vertices $s+1,\ldots,i-1$ 
(and also sampled $b_{s+1}, \ldots, b_{i-1}$), and want to fill in the edges from $S$ to vertex $i$.
Conditioned on $b_{s+1:i-1}$, these edges have the following distribution under $P_0$:
\begin{itemize}
\item For each $j=1,\ldots,s$, set $A_{ji} = 1$ with probability $q$.
\item Additionally, sample $j_0 \in \{1,\ldots,m\}$ with probability 
      $\pi_{j_0} = \frac{m-1-N_i(j_0)}{m^2-m-(i-s-1)}$. 
      If $j_0 \in [s]$ then set $A_{j_0i} = 1$ as well.
\end{itemize}
Here $N_i(j)$ counts the number of $b_l$ with $b_l = j$ and $s+1 \leq l < i$. 
This expression for $\pi_{j_0}$ comes from the fact that among the $m^2-m-(i-s-1)$ pairs 
$(a,b)$ that have not yet been used, there are $m-1$ pairs $(a,b)$ with 
$a \neq 1$ and $b = j_0$, but some might have been used up already from the samples 
$b_{s+1}, \ldots, b_{i-1}$.

Note that $A_{1:s,i}$ is exactly sampled from a perturbed Bernoulli distribution 
$\PB(q, \sigma)$, where $\sigma({\{j\}}) = \pi_j$, 
$\sigma({\emptyset}) = 1-\sum_{j=1}^s \pi_j$, and $\sigma(J) = 0$ for $|J| \geq 2$.
On the other hand, under $P_1$, the $A_{1:s,i}$ are instead sampled independently from 
$\Ber(\frac{1}{2})$. This is equivalent to a $\PB(q, \tau)$ distribution 
with $\tau(J) = (\frac{1-2q}{2-2q})^{|J|}(\frac{1}{2-2q})^{s-|J|}$ (see comment 
at end of Section~\ref{sec:pb}). We can therefore apply 
Proposition~\ref{prop:kl-multi-2} regarding the KL divergence of perturbed Bernoulli 
distributions to obtain the following bound on $\DKL(P_0(A_{1:s,i}), P_1(A_{1:s,i}))$:
\begin{corollary}
For $P_0$ and $P_1$ as above, we have
\begin{align}
\nonumber \lefteqn{\DKL(P_0(A_{1:s,i}), P_1(A_{1:s,i}) \mid b_{1:i-1}, s)} \\
\nonumber &\leq (2-2q)^s \Big(\Big(\frac{1-q}{q}\Big)^2 \sum_{j=1}^s \Big(\frac{1-2q}{2-2q} - \pi_j\Big)^2 + \sum_{|J| \geq 2} \Big(\frac{(1-q)(1-2q)}{q(2-2q)}\Big)^{2|J|}\Big) \\
 &= \big(\frac{m}{m-1}\big)^s \Big(\big(\frac{m}{m-2}\big)^2\sum_{j=1}^s (1/m - \pi_j)^2 + \sum_{|J| \geq 2} (m-2)^{-2|J|}\Big).
\label{eq:kl-cor-1}
\end{align}
\end{corollary}
In the first line, 
the $(2-2q)^s$ term comes from the fact that $\tau({\emptyset}) = (\frac{1}{2-2q})^s$; the 
first sum comes from comparing $S({\{j\}})$ and $T({\{j\}})$ (sets of size $1$), 
while the second sum comes from summing over $|J| \geq 2$ and noting that $S(J) = 0$ 
for such sets. The second line is simply substituting $q = \frac{1}{2} - \frac{1}{2m-2}$.

Provided that $s \leq m-6$ we can check that 
$(\frac{m}{m-1})^s(\frac{m}{m-2})^2 \leq 3$, and 
$\sum_{|J| \geq 2} (m-2)^{-2|J|} = (1+\frac{1}{(m-2)^2})^s - \frac{s}{(m-2)^2} - 1 \leq \frac{s^2}{(m-2)^4}$. Therefore, \eqref{eq:kl-cor-1} yields
\begin{equation}
\label{eq:kl-bound-local}
\DKL(P_0(A_{1:s,i}), P_1(A_{1:s,i}) \mid b_{1:i-1}, s) \leq 3\frac{s^2}{(m-2)^4} + 3\sum_{j=1}^s (\pi_j(b_{1:i-1}) - 1/m)^2.
\end{equation}

\paragraph{Chaining the KL divergence.}
Having bounded the KL divergence of each local decision, we would like to obtain a global 
bound on the difference between $P_0$ and $P_1$. We can do this with the following inequality, 
which follows from the chain rule for KL divergence (all expectations are with respect to $P_0$):

\begin{equation}
\label{eq:kl-chain}
\DKL(P_0(A), P_1(A)) \leq \bE_{s}\bigg[ \sum_{i=s+1}^n \bE_{b_{1:i-1}}\left[ \DKL(P_0(A_{1:s,i}), P_1(A_{1:s,i}) \mid b_{1:i-1}, s) \right] \bigg].
\end{equation}
Plugging \eqref{eq:kl-bound-local} into \eqref{eq:kl-chain}, we obtain
\begin{equation}
\label{eq:kl-bound-global-0}
\DKL(P_0(A), P_1(A)) \leq \bE_{s}\bigg[ \frac{3s^2n}{(m-2)^4} + 3\sum_{i=s+1}^n \sum_{j=1}^s \bE_{b_{1:i-1}}\left[(\pi_j(b_{1:i-1}) - 1/m)^2\right] \bigg].
\end{equation}
To bound \eqref{eq:kl-bound-global-0} we need to analyze the mean and variance of $\pi_j$. 
By symmetry, $\bE[\pi_j(b_{1:i-1})] = \pi_j(b_{1:s}) = \frac{1}{m}$. Also, 
$\Var[\pi_j(b_{1:i-1})] = \frac{1}{(m^2-m-(i-s-1))^2} \Var[\sum_{l=s+1}^{i-1} \bI[b_l=j]]$. 
We have $(m^2-m-(i-s-1))^2 \geq (m^2-2n)^2 \geq \frac{1}{2}m^4$ assuming $m^2 \geq 7n$. 
Therefore, $\Var[\pi_j]$ is bounded above as
\begin{align}
\Var[\pi_j(b_{1:i-1})] 
 &\leq \frac{2}{m^4} \Var\Big[\sum_{l=s+1}^{i-1} \bI[b_l=j]\Big] \\
 &\stackrel{(i)}{\leq} \frac{2}{m^4} \sum_{l=s+1}^{i-1} \Var[\bI[b_l=j]] \\
 &\stackrel{(ii)}{\leq} \frac{2(i-s-1)}{m^5} \leq \frac{2n}{m^5}.
\end{align}
Here (i) is because the events $\bI[b_l=j]$ are negatively correlated 
(since we sample without replacement), and (ii) is because $\bI[b_l=j]$ is 
$\Ber(\frac{1}{m})$ which has variance less than $\frac{1}{m}$. 
We thus have 
$\bE_{b_{1:i-1}}[(\pi_j(b_{1:i-1})-1/m)^2] = \Var_{b_{1:i-1}}[\pi_j(b_{1:i-1})] \leq \frac{2n}{m^5}$.

Now, plugging back into \eqref{eq:kl-bound-global-0} and using the fact that $\bE[s] = n/m$, 
$\Var[s] \leq n/m$, we obtain
\begin{align}
\DKL(P_0(A), P_1(A)) 
 &\leq \bE_s\left[\frac{3s^2n}{(m-2)^4} + \frac{6sn^2}{m^5}\right] \\
 &\leq \frac{3n}{(m-2)^4}((n/m)^2 + n/m) + \frac{6n^3}{m^6} \leq \oo\Big(\frac{n^3}{m^6}\Big).
\end{align}
By Pinsker's inequality, this implies that 
$\DTV(P_0(A), P_1(A)) \leq \sqrt{\frac{1}{2}\DKL(P_0(A), P_1(A))} = \oo(n^{1.5}/m^3)$.

\paragraph{Obtaining a recovery lower bound.}
We have just concluded that the total variation distance between $P_0$ and $P_1$ is small---at 
most $\oo(n^{1.5}/m^3)$. On the other hand, under the null distribution $P_0$ it is clear 
that $S$ cannot be recovered with probability greater than $\frac{1}{2}$ (because there 
are two completely symmetric choices for $S$). The total variation distance upper-bounds 
the probability that \emph{any} test can distinguish between $P_0$ and $P_1$. In particular, 
if it was possible to recover $S$ with probability much greater than 
$\frac{1}{2} + \Theta(n^{1.5}/m^3)$, then we could use this as a test, which would contradict 
the total variation bound. So, there is no way to recover $S$ with probability 
greater than $\frac{1}{2} + \oo(n^{1.5}/m^3)$. 

\section{Lower Bound: Full Argument}
\label{sec:lower-full}

We now give the argument for the full lower bound, showing that the probability of 
recovering $S$ (and more generally, the Jaccard similarity with $S$) goes to $0$, 
rather than to $\frac{1}{2}$ as above.
In the previous section, our construction involved a null distribution $P_0$ where 
every vertex belonged to $2$ cliques. In order for the recovery probability to go 
to $0$, we need every vertex to belong to $k$ cliques, where $k \to \infty$ (we will 
eventually take $k = \Theta(\sqrt{n}/s)$.

The analysis is nearly identical to Section~\ref{sec:lower}, except that the probability 
distribution over each column $A_{1:s,i}$ becomes much more complicated. Before it consisted of 
a Bernoulli distribution perturbed in at most one element, whereas now each column could 
(with some low probability) be perturbed in as many as $k-1$ places, and the overall 
distribution over perturbations ends up being quite complex. Fortunately, it is still 
in the family of perturbed Bernoulli distributions, and so we can still apply 
Proposition~\ref{prop:pb} to obtain bounds on the KL divergence.

To start, we first need to generalize the construction from Section~\ref{sec:lower} from 
$2$ sets of cliques to $k$ sets of cliques. 
The previous construction sampled $(a_i, b_i)$ from an $m \times m$ grid, 
and created cliques corresponding to the rows and columns of that grid. If we wanted 
a third set of cliques, we could also consider diagonals of the grid. More generally, 
we can associate $[m] \times [m]$ with the two-dimensional vector space $\bF_m^2$ (assuming 
$m$ is prime), and construct a clique for every line of slope $0, 1, \ldots, k-1$. 

\paragraph{The null distribution $P_0$.} 
More formally, the null distribution $P_0$ is constructed as follows:
\begin{itemize}
\item For $i = 1, \ldots, n$, sample $(a_i, b_i)$ without replacement 
      from $[m] \times [m]$.
\item Define the relation $(a,b) \rel (a',b')$ if $a - a' \equiv r (b - b') \pmod{m}$ for 
      some $r \in \{0, \ldots, k-1\}$.
\item Generate $A$ by setting 
      $A_{ij} = \left\{\begin{array}{ccl} 1 & : & (a_i,b_i) \rel (a_j,b_j) \\ \Ber(q) & : & \text{else.} \end{array} \right.$
\end{itemize}
Here we will let $q = \frac{1}{2} - \frac{k-1}{2(m-k+1)}$.

\paragraph{The semi-random instance $P_1$.}
As in the previous section, we will construct $P_1$ via a coupling with $P_0$. 
The idea will be to pick a random line $L_{r,h} = \{(a,b) \mid a - br \equiv h \pmod{m}\}$ 
as corresponding to the true planted clique. For this line, the outgoing edges will 
all be sampled from $\Ber(\frac{1}{2})$ rather than according to the null distribution. 
Specifically, $P_1$ is sampled as follows, analogously to Section~\ref{sec:lower}:
\begin{enumerate}
\item[0.] Sample $r^*$ uniformly from $\{0,\ldots.k-1\}$ and $h^*$ uniformly from $\{0,\ldots,m-1\}$.
\item[1.] Sample $s \sim \HG(n, m, m^2)$ and let $S$ be a uniformly random subset of 
      $[n]$ of size $s$. For $i \in S$, sample $(a_i, b_i)$ uniformly from 
      $L_{r^*,h^*}$ without replacement.
\item[$2'$.] For each $i\not\in S$ and $j \in S$, draw an edge between $j$ and $i$ with probability 
          $\frac{1}{2}$.
\item[3.] Fill in the remaining edges according to the conditional distribution under $P_0$.
\end{enumerate}
We refer to the second step as $2'$ because it is analogous to step $2'$ in the 
previous section. This semi-random instance is almost identical to the null distribution $P_0$, 
except that under the null distribution step $2'$ is replaced with the following step $2$:
\begin{enumerate}
\item [2.] For each $i\not\in S$, sample $(a_i,b_i)$ without replacement from 
           $\bF_m^2 \backslash L_{r^*,h^*}$. Then for $j \in S$, if $(a_j,b_j) \rel (a_i,b_i)$ draw an edge between $j$ and $i$; otherwise draw an edge between $j$ and $i$ with probability $q$.
\end{enumerate}
Note that step $2$ again corresponds to a perturbed Bernoulli distribution, where the 
perturbations correspond to the edges $A_{ji}$ with $(a_j,b_j) \rel (a_i,b_i)$; we analyze 
this in more detail below.

\paragraph{Bounding $\DKL(P_0, P_1)$.}
Without loss of generality assume that $S = \{1,\ldots,s\}$. 
As before, the main task is bounding the conditional KL divergence 
$\DKL(P_0(A_{1:s,i}), P_1(A_{1:s,i}) \mid a_{1:i-1}, b_{1:i-1}, s)$ for the edges 
between $S$ and some $i \not\in S$.

Under the null distribution $P_0$, the conditional distribution of $A_{1:s,i}$ is a 
perturbed Bernoulli distribution: 
first we sample $Y_{1:s} \sim \Ber(q)$, then sample $(a_i, b_i)$ uniformly from 
$\bF_m^2 \backslash (L_{r^*,h^*} \cup \{(a_{l}, b_{l})\}_{l=s+1}^{i-1})$ and 
set $A_{ji} = 1$ if either $Y_j = 1$ or $j \in J(a_i, b_i)$, where 
$J(a,b) = \{j \in [s] \mid (a_j,b_j) \rel (a,b)\}$.

This is a more complicated distribution from before, but 
Proposition~\ref{prop:kl-multi-2} is still flexible enough to analyze it. 
Let $N_i(j)$ be the number of $(a_{l}, b_{l})$ with $s+1 \leq l < i$ such that 
$(a_{l}, b_l) \rel (a_j, b_j)$. Then
$A_{1:s,i} \sim \PB(q, \sigma)$, where $\sigma$ does not have a closed form, but 
for a single element $j$ we have 
\begin{equation}
\label{eq:S-comp}
S({\{j\}}) = \sum_{J \supseteq \{j\}} \sigma({J}) = \frac{(k-1)(m-1) - N_i(j)}{m^2 - m - (i-s-1)}.
\end{equation}
This is because there are initially $(k-1)(m-1)$ opportunities for $i$ and $j$ to lie on the 
same line (note that $L_{r^*,h^*}$ is already used up), and $N_i(j)$ counts the number 
of these opportunities that are already taken up by previous $(a_l, b_l)$.

On the other hand, for $|J| = 2$, first note that 
$S(J) \leq \frac{(k-1)(k-2)}{m(m-1)-n} \leq \frac{2k^2}{m^2}$ (assuming $n \leq m(m-1)/2$). 
This is because for any two points $(a,b), (a',b') \in L_{r^*, h^*}$, any pair of non-parallel 
lines through these points has a single point of intersection, and so the probability of 
containing the set $J = \{(a,b), (a',b')\}$ is at most 
$(k-1)(k-2)$ (the number of pairs of non-parallel lines) divided by 
$m(m-1)-n$ (a lower bound on the number of points in $\bF_m^2 \backslash L_{r^*,h^*}$ 
that have not yet been sampled). Since $S(J)$ decreases monotonically in $J$, this 
means that $S(J) \leq \frac{2k^2}{m^2}$ for all $|J| \geq 2$ as well.

We can now apply Proposition~\ref{prop:kl-multi-2}. As before, under $P_1$, $A_{1:s,i}$ 
comes from $\Ber(\frac{1}{2}) = \PB(q, \tau)$, with 
$\tau(J) = (\frac{1-2q}{2-2q})^{|J|}(\frac{1}{2-2q})^{s-|J|}$. Assuming $k \leq m/4$ and 
$2s+8 \leq m/k$, we get (see Appendix~\ref{sec:pb-full-details} for details)
\begin{align}
\nonumber \lefteqn{\DKL(P_0(A_{1:s,i}), P_1(A_{1:s,i}) \mid a_{1:i-1}, b_{1:i-1}, s)} \\
 &\phantom{+} \leq 3\sum_{j=1}^s (S(\{j\}) - T(\{j\}))^2 + 6\sum_{|J| \geq 2} \Big(\frac{m}{m-2k+2}\Big)^{2|J|} (S(J)^2 + T(J)^2).
\label{eq:pb-full}
\end{align}
Note that $T(J) = (\frac{k-1}{m})^{|J|}$. For $|J| = 1$, 
this together with \eqref{eq:S-comp} yields, for the first sum,
\begin{align}
\sum_{j=1}^s (S({\{j\}}) - T({\{j\}}))^2 
 &= \sum_{j=1}^s \Big(\frac{(k-1)(m-1) - N_i(j)}{m^2-m-(i-s-1)} - \frac{k-1}{m}\Big)^2.
\end{align}
Next, plugging in $T(J) = (\frac{k-1}{m})^{|J|}$ for larger $J$ yields 
\begin{align}
\sum_{|J| \geq 2} \Big(\frac{m}{m-2k+2}\Big)^{2|J|} T(J)^2
 &= \sum_{|J| \geq 2} \Big(\frac{m}{m-2k+2}\frac{k-1}{m}\Big)^{2|J|} \\
 &= \Big(1 + \big(\frac{k-1}{m-2k+2}\big)^2\Big)^s - \big(\frac{k-1}{m-2k+2}s\big)^2 - 1 \\
 &\leq \Big(\frac{s(k-1)^2}{(m-2k+2)^2}\Big)^2 \leq \frac{2k^4s^2}{m^4}
\end{align}
assuming that $s(k-1) \leq m-2k+2$.
Finally, using the previous observation $S(J) \leq \frac{2k^2}{m^2}$, we have
\begin{align}
\sum_{|J| \geq 2} S(J)^2 
 &\leq \frac{2k^2}{m^2} \sum_{|J| \geq 2} S(J).
\end{align}
Putting these together, we get:
\begin{corollary}
Assume that $k \leq \frac{m}{4}$ and that $s \leq \frac{m}{2k} - 4$. Then we have
\begin{align}
\nonumber \lefteqn{\DKL(P_0(A_{1:s,i}), P_1(A_{1:s,i}) \mid a_{1:i-1}, b_{1:i-1}, s)} \\
 &\phantom{++} \leq 3 \sum_{j=1}^s \Big(\frac{(k-1)(m-1) - N_i(j)}{m^2-m-(i-s-1)} - \frac{k-1}{m}\Big)^2 + \frac{12k^4s^2}{m^4} + \frac{12k^2}{m^2} \sum_{|J| \geq 2} S(J).
\end{align}
\end{corollary}
\paragraph{Chaining the KL divergence.}
Chaining the KL divergence in the same way as before, we obtain for any $s \leq \frac{m}{2k}-4$,
\begin{align}
\nonumber \lefteqn{\DKL(P_0(A), P_1(A) \mid s)} \\
 &\leq \sum_{i=s+1}^n \bE_{a_{1:i-1}, b_{1:i-1}}[\DKL(P_0(A_{1:s,i}), P_1(A_{1:s,i}) \mid a_{1:i-1}, b_{1:i-1}, s)] \\
 &\leq \frac{12k^4s^2n}{m^4} + \frac{12k^2}{m^2} \sum_{i=1}^n \sum_{|J| \geq 2} \bE[S(J) \mid s] + 3\sum_{i=1}^n \sum_{j=1}^s \bE\bigg[\Big(\frac{(k-1)(m-1) - N_i(j)}{m^2-m-(i-s-1)} - \frac{k-1}{m}\Big)^2 \mid s\bigg].
\end{align}
Now, for the first sum we have
\begin{align}
\sum_{|J| \geq 2} S(J) &= \sum_{|J| \geq 2} \sum_{J' \supseteq J} \sigma({J'}) \\
 &\leq \sum_{J'} \sum_{J \subseteq J', |J| \geq 2} \sigma({J'}) \\
 &\leq \sum_{J'} (2^{|J'|} - |J'| - 1) \sigma({J'}),
\label{eq:I-pre}
\end{align}
which is the expected value of $2^I-I-1$, where $I$ is the number of points of 
intersection of $S$ with the $k-1$ lines going through $(a_i,b_i)$. 
Since each of these lines intersects $L_{r^*,h^*}$ exactly once, 
and $S$ is a uniformly random subset of $s$ out of $m$ elements of 
$L_{r^*,h^*}$, $I$ is hypergeometrically distributed--$I \sim \HG(k-1,s,m)$--and 
\eqref{eq:I-pre} is equal to $\bE_{I \sim \HG(k-1,s,m)}[2^I-I-1] \leq \frac{4k^2s^2}{m^2}$ 
as long as $s \leq m/2k$ (see Appendix~\ref{sec:hg-details} for justification).

For the second sum, note that $\frac{k-1}{m}$ is the mean of the random 
variable $\frac{(k-1)(m-1)-N_i(j)}{m^2-m-(i-s-1)}$, and so the calculation comes 
down to computing the variance of $\Var[N_i(j)]/(m^2-m-(i-s-1))^2 \leq \frac{4}{m^4} \Var[N_j]$ 
assuming that $n+m \leq m^2/2$. 
We note that $N_i(j)$ is a sum of $k-1$ (dependent) hypergeometric distributions 
each distributed as $\HG(i-s-1, (k-1)(m-1), m(m-1))$ and so has variance at most $k^2 \cdot (i-s-1)\frac{k-1}{m} \leq \frac{nk^3}{m}$; the overall sum is then at most $\frac{4k^3sn^2}{m^5}$. 
Putting these together, we obtain:
\begin{proposition}
\label{prop:lower}
If $k \leq \frac{m}{4}$, $s \leq \frac{m}{2k}-4$, and $n+m \leq \frac{m^2}{2}$, then we have 
\begin{equation}
\DKL(P_0(A), P_1(A) \mid s) \leq \oo\Big(\frac{k^4s^2n}{m^4} + \frac{k^3sn^2}{m^5}\Big).
\end{equation}
\end{proposition}

\paragraph{Proof of Theorem~\ref{thm:main-lower}.}
We now have all we need to prove our main result. 
We will take our distribution over semi-random 
instances to be 
$P_1(A \mid \frac{n}{2m} \leq s \leq \frac{2n}{m})$. Assuming that $n \leq \frac{m^2}{4k} - 2m$, 
we have $\frac{2n}{m} \leq \frac{m}{2k}-4$ and so we can apply Proposition~\ref{prop:lower}. 
Therefore, the KL divergence between $P_1$ and $P_0$ 
(conditioned on $s \in [\frac{n}{2m}, \frac{2n}{m}]$) 
is at most 
$\oo(\bE[\frac{k^4n}{m^4}s^2 + \frac{k^3n^2}{m^5}s \mid \frac{n}{2m} \leq s \leq \frac{2n}{m}]) = \oo(k^4n^3/m^6)$. Therefore, by Pinsker's inequality, 
$\DTV(P_0(A), P_1(A) \mid \frac{n}{2m} \leq s \leq \frac{2n}{m}) = \oo(k^2n^{1.5}/m^3)$.

On the other hand, under $P_0$, $s$ is a hypergeometric random variable 
$\HG(n,m,m^2)$, and so the probability that $s \in [\frac{n}{2m}, \frac{2n}{m}]$ 
is very high: at least $1 - 2e^{-n/8m}$ by \citet{hoeffding1963probability}. Therefore, 
the unconditional distribution of $A$ is close to that conditioned on 
$s \in [\frac{n}{2m}, \frac{2n}{m}]$: 
$\DTV(P_0(A), P_0(A \mid \frac{n}{2m} \leq s \leq \frac{2n}{m})) \leq 2e^{-n/8m}$. 
Together with the previous bound, this yields 
$\DTV(P_1(A \mid \frac{n}{2m} \leq s \leq \frac{2n}{m}), P_0(A)) \leq \oo(k^2n^{1.5}/m^3 + e^{-n/8m})$.

Now, under $P_0$, the expected Jaccard similarity is clearly $\oo(1/k)$ (because there are $k$ 
identical cliques to which $v$ belongs, so one cannot align with the true clique by 
more than $\Theta(1/k)$ on average). On the other hand, the Jaccard similarity is bounded 
between $0$ and $1$, so the total variational bound implies that
\begin{equation}
\bE[J(S, \hat{S}(A,v))] \leq \oo\Big(\frac{1}{k} + \frac{k^2n^{1.5}}{m^3} + e^{-n/8m}\Big)
\end{equation}
as long as $n \leq \frac{m^2}{4k} - 2m$ and $k \leq \frac{m}{4}$. 
We will take $k = m/\sqrt{n}$, which satisfies the conditions as long as 
$16 \leq n \leq m^2/64$.
This yields a bound of $\oo(\sqrt{n}/m + e^{-n/8m})$. Since the minimum clique 
size $s_0$ is $\frac{n}{2m}$, this is the same as 
$\oo(s_0/\sqrt{n} + e^{-s_0/4})$, which is small as long as 
$s_0 \geq 8\log(n)$. We therefore obtain the bound $\oo\big(\frac{s+\log(n)}{\sqrt{n}}\big)$. 
The condition $n \leq m^2/64$ necessitates $s \leq \sqrt{n}/2$, but this holds whenever 
the error bound is non-vacuous. We therefore get 
$\bE[J(S, \hat{S}(A,v))] = \oo\big(\frac{s+\log(n)}{\sqrt{n}}\big)$, as was to be shown. 

\section{Upper Bound}
\label{sec:upper}

We next turn to the upper bound (Theorem~\ref{thm:main-upper}). The crux is the following 
lemma showing that the planted clique is nearly disjoint from all other large cliques:
\begin{proposition}
\label{prop:union-bound}
Let $S$ be a planted clique of size $s$ in a subgraph of size $n$ under the semi-random model. 
Then, with probability at least $1-\frac{2s}{n^2}$, any other clique 
$S'$ of size at least $s$ satisfies $|S \cap S'| < 3\log_2(n)$.
\end{proposition}
Now, call a clique \emph{good} if it has size at least $s$, and its intersection 
with any other clique of size at least $s$ is at most $3\log_2(n)$. We have just seen 
that the planted clique $S$ is good with probability $1-o(1)$. Moreover, if 
$s \geq 3\sqrt{n\log_2(n)}$, then there must be less than $\frac{2n}{s}$ good cliques. 
To see this, note that by the principle of inclusion-exclusion, the union of $m$ good 
cliques has size at least $ms - 3\binom{m}{2}\log_2(n) > ms - \frac{3}{2}m^2\log_2(n)$, and so 
we must have $m(s-\frac{3}{2}m\log_2(n)) < n$. If we take $m = \frac{2n}{s}$ then we obtain
$\frac{3}{2}m\log_2(n) < \frac{s}{2}$, and hence $m(s-\frac{3}{2}m\log_2(n)) \geq ms/2 = n$, which is a 
contradiction. This shows that we indeed have $m < \frac{2n}{s}$.

Now, since there are at most $\frac{2n}{s}$ good cliques, the total number of vertices 
in $S$ that intersect with any other good clique is at most 
$\frac{6n\log_2(n)}{s}$, and so the fraction of such vertices in $S$ is at most 
$\frac{6n\log_2(n)}{s^2}$. This yields the following recovery algorithm: given $A$ and $v$, 
if $v$ lies in a unique good clique then output that clique as $S$; otherwise, output the 
empty set. With probability $1 - \frac{2s}{n^2} - \frac{6n\log_2(n)}{s^2}$, this gives us 
exact recovery of the planted clique $S$, which completes the first part of 
Theorem~\ref{thm:main-upper}.

For the second part, we invoke Corollary 9.3 of \citet{charikar2017learning}. While 
their result is more general, in our context it specializes to the following:
\begin{theorem}[\citet{charikar2017learning}]
\label{thm:charikar}
Let $A$ be a graph drawn from the semi-random model with a planted clique of size $s$. 
Then there is a polynomial time algorithm which, with probability $1 - \exp(-\Omega(s))$, 
outputs sets $\hat{S}_1, \ldots, \hat{S}_m$ with $m \leq \frac{4n}{s}$, such that 
$\min_{j=1}^m |S \triangle \hat{S}_j| = \oo\p{(n/s)^2\log(n)}$.
\end{theorem}
Now for a large enough constant $C$, if $s \geq C \cdot n^{2/3}\log^{1/3}(n)$ then 
the bound in Theorem~\ref{thm:charikar} translates to 
$|S \triangle \hat{S}_j| \leq \frac{s}{8}$. Then, any vertex $i \in S$ will be connected 
to at least $\frac{7s}{8}$ elements in $\hat{S}_j$, while with probability 
$1 - n\exp(-\Omega(s))$, no vertex not in $S$ will be connected to more than 
$\frac{3s}{4}$ elements in $\hat{S}_j$. We can thus define $\tilde{S}_j$ to be the set of 
vertices that are connected to at least $\frac{7s}{8}$ elements in $\hat{S}_j$, and with 
high probability one of the $\tilde{S}_j$ will be the planted clique $S$. 

To finish, we remove any $\tilde{S}_j$ that is not a clique of size at least $s$, and then 
also any $\tilde{S}_j$ that has intersection greater than $3\log_2(n)$ with 
any of the other remaining $\tilde{S}_{j'}$. 
Given a vertex $v$, we return the $\tilde{S}_j$ that it 
belongs to (if it exists and is unique) and otherwise return the empty set. By the same logic 
as before, this outputs $S$ with probability at least 
$1 - \frac{2s}{n^2} - \frac{12n\log_2(n)}{s^2} - (n+1)\exp(-\Omega(s)) = 1-o(1)$. 

This completes the proof of Theorem~\ref{thm:main-upper}.
We remark that Proposition~\ref{prop:union-bound} remains true under 
\citeauthor{feige2001heuristics}'s model, and hence the information-theoretic 
part of Theorem~\ref{thm:main-upper}
holds in that model as well.

\appendix

\section{Proof of Proposition~\ref{prop:kl-multi-2}}
\label{sec:kl-multi-2-proof}

By summing over the different possible samples from $\sigma$, we can calculate 
$P_X$ as 
\begin{align}
P_{X}(x_{1:s}) &= \sum_{J} \sigma(J) \bI[x_J=1] \prod_{j \not\in J} q^{x_j}(1-q)^{1-x_j} \\
 &= \bigg(\prod_{j=1}^s q^{x_j}(1-q)^{1-x_j}\bigg) \bigg(\sum_{J} \sigma(J) \bI[x_J=1]/q^{|J|}\bigg) \\
 &\stackrel{(i)}{=} \bigg(\prod_{j=1}^s q^{x_j}(1-q)^{1-x_j}\bigg) \bigg(\sum_{J} S(J) \prod_{j \in J} (-1 + \bI[x_j=1]/q)\bigg).
\label{eq:y-2}
\end{align}
To justify (i), for any $J_0$ consider all occurences of the term 
$\bI[x_{J_0} = 1]/q^{|J_0|} = \prod_{j \in {J_0}} (\bI[x_j=1]/q)$. 
For each $J \supseteq J_0$, this term will occur with coefficient 
$(-1)^{|J \backslash J_0|} S(J) = (-1)^{|J \backslash J_0|} \sum_{J' \supseteq J} \sigma({J'})$. 
Therefore, the overall coefficient is 
$\sum_{J' \supseteq J_0} \big(\sigma({J'}) \sum_{J : J_0 \subseteq J \subseteq J'} (-1)^{|J \backslash J_0|}\big) = \sigma({J_0})$, since the inner sum is zero unless $J' = J_0$.

Motivated by this, we define $y_J(x) = S(J) \prod_{j \in J} (-1 + \bI[x_j=1]/q)$. An analogous 
derivation to \eqref{eq:y-2} holds for $P_{X'}$, and we correspondingly define 
$z_J(x) = T(J) \prod_{j \in J} (-1 + \bI[x_j=1]/q)$.

By Lemma 2.7 of \citet{tsybakov2009introduction}, we can upper bound KL divergence by 
$\chi^2$-divergence:
\begin{align}
\DKL(P_X, P_{X'}) 
 &\leq D_{\chi^2}(P_X, P_{X'}) \\
 &= \sum_{x \in \{0,1\}^n} \frac{(P_X(x) - P_{X'}(x))^2}{P_{X'}(x)} \\
 &= \sum_{x \in \{0,1\}^n} \bigg(\prod_{j=1}^s q^{x_j}(1-q)^{1-x_j}\bigg) \frac{\p{\sum_{J} \big(z_J(x) - y_J(x)\big)}^2}{\sum_J \tau(J) \bI[x_J=1]/q^{|J|}}.
\end{align}
We can always bound the denominator by simply $\tau({\emptyset})$ (since the rest of 
the terms in the sum are non-negative). In addition, we can treat the 
sum over $x \in \{0,1\}^s$ as an expectation with respect 
to a $\Ber(q)$ distribution. Together, these yield
\begin{align}
\label{eq:kl-multi-0-2}
\DKL(P_X, P_{X'}) 
 &\leq \frac{1}{\tau({\emptyset})} \cdot \bE_{x_{1:s} \sim \Ber(q)}\left[\Bigg(\sum_{J} \big(z_J(x) - y_J(x)\big)\Bigg)^2\right].
\end{align}
Now note that $\bE_{x \sim \Ber(q)}[z_J(x)] = \bE_{x \sim \Ber(q)}[y_J(x)] = 0$ for all $J$.
Also, for $J \neq J'$, we have (by independence of the $x_j$) 
\begin{equation}
\bE[y_J(x)y_{J'}(x)] = S(J)S(J') \prod_{j \in J \cap J'} \bE[(-1 + \bI[x_j=1]/q)^2] \prod_{j \in J \triangle J'} \bE[(-1 + \bI[x_j=1]/q)] = 0,
\end{equation}
since all the terms in the latter product are zero. By analogous arguments, 
$\bE[z_J(x)y_{J'}(x)] = 0$ and $\bE[z_J(x)z_{J'}(x)] = 0$. 
%
Together these 
imply that most of the terms in \eqref{eq:kl-multi-0-2} are $0$; indeed, we have 
\begin{align}
\bE_{x}\Bigg[\Bigg(\sum_{J} \big(z_J(x) - y_J(x)\big))\Bigg)^2\Bigg]
 &= \bE_{x}\Bigg[\sum_{J} \big(z_J(x) - y_J(x)\big)^2\Bigg].
\end{align}
Now, we have 
$(z_J(x) - y_J(x))^2 = (S(J) - T(J))^2 \prod_{j \in J} (-1 + \bI[x_j=1]/q)^2 \leq (S(J) - T(J))^2 \Big(\frac{1-q}{q}\Big)^{2|J|}$. 
We thus obtain 
\begin{align}
\label{eq:kl-multi-1-2}
\DKL(P_X, P_{X'}) 
 &\leq \frac{1}{\tau({\emptyset})} \sum_{J} \Big(\frac{1-q}{q}\Big)^{2|J|} (S(J) - T(J))^2,
\end{align}
as was to be shown.

\section{Proof of Equation (\ref{eq:kl-chain})}
\label{sec:kl-chain-proof}

We make use of the chain rule for KL divergence, which 
says that given distributions $P_0(X_{1:N})$ and $P_1(X_{1:N})$, 
we have $\DKL(P_0(X_{1:N}), P_1(X_{1:N})) = \sum_{i=1}^N \bE[\DKL(P_0(X_i), P_1(X_i) \mid X_{1:i-1})]$ (all expectations are with respect to $P_0$).

In our case, we want to bound $\DKL(P_0(A), P_1(A))$. We will add in the 
auxiliary variables $b_{1:n}$ and $s$, which will only increase the KL divergence, 
and then apply the chain rule. For short-hand, we use $\DKL(X \mid Y)$ to denote 
$\DKL(P_0(X \mid Y), P_1(X \mid Y))$. We have:
\begin{align}
\DKL(P_0(A), P_1(A))
 &\leq \DKL(P_0(A, b, s), P_1(A, b, s)) \\
\notag
 &= \DKL(s, A_{1:s,1:s}, b_{1:s}) + \bE_s\bigg[\sum_{i=s+1}^n \bE[\DKL(A_{1:s,i}, b_i \mid s, A_{1:s,1:i-1}, b_{1:i-1})]\bigg] \\
 &\phantom{++} + \bE[\DKL(A_{s+1:n,s+1:n} \mid s, A_{1:s,1:n}, b_{1:n})] \\
 &= \bE_s\bigg[\sum_{i=s+1}^n \bE[\DKL(A_{1:s,i}, b_i \mid s, A_{1:s,1:i-1}, b_{1:i-1})] \bigg],
\label{eq:kl-1}
\end{align}
where the final equality is because all of the other conditional distributions are identical 
under $P_0$ and $P_1$.

Furthermore, $A_{1:s,i}, b_i$ are independent of $A_{1:s,1:i-1}$ conditioned on 
$s$ and $b_{1:i-1}$, so 
\begin{align}
\DKL(A_{1:s,i}, b_i \mid s, A_{1:s,1:i-1}, b_{1:i-1}) 
 &= \DKL(A_{1:s,i}, b_i \mid s, b_{1:i-1}) \\
 &= \DKL(A_{1:s,i} \mid s, b_{1:i-1}) + \bE[\DKL(b_i \mid s, b_{1:i-1}, A_{1:s,i})] \\
 &= \DKL(A_{1:s,i} \mid s, b_{1:i-1}).
\label{eq:kl-2}
\end{align}
Here again the final equality is because $b_i$ has an identical conditional distribution 
under $P_0$ and $P_1$.

Plugging \eqref{eq:kl-2} into \eqref{eq:kl-1}, we obtain 
\begin{align}
\DKL(P_0(A), P_1(A))
 &\leq \bE_s\bigg[\sum_{i=s+1}^n \bE_{b_{1:i-1}}[\DKL(A_{1:s,i} \mid s, b_{1:i-1})]\bigg],
\end{align}
which is exactly the statement of \eqref{eq:kl-chain}.

\section{Proof of Proposition~\ref{prop:union-bound}}
\label{sec:union-bound-proof}

Take any candidate clique $S'$, which we can assume has size exactly $s$ (since any larger 
$S''$ would contain a clique $S'$ of size $s$).
For any such $S'$, all of the edges between $S$ and $S'$ must be 
present, which occurs with probability $(1/2)^{l(s-l)}$, where $l = |S \cap S'|$.
On the other hand, there are $\binom{s}{l}\binom{n-s}{s-l}$ sets of size $s$ with 
intersection $l$. Union bounding over all $l \geq l_0$, the probability that 
there is any clique with intersection greater than $l_0 = 3\log_2(n)$ is at most
\begin{align}
\sum_{l_0 \leq l < s} \binom{s}{l}\binom{n-s}{s-l}2^{-l(s-l)}
 &\leq \sum_{l_0 \leq l < s} \binom{s}{l} \p{\frac{n-s}{2^l}}^{s-l} \\
 &\leq \sum_{l_0 \leq l < s} \binom{s}{l} (1/n^2)^{s-l} \\
 &\leq (1+1/n^2)^{s} - 1 \leq 1 + \frac{2s}{n^2},
\end{align}
where the final inequality holds because $(1+1/n^2)^s \leq \exp(s/n^2) \leq 1+2s/n^2$ 
since $s/n^2 \leq 1$.

\section{Details of Equation \ref{eq:pb-full}}
\label{sec:pb-full-details}

We have 
\begin{align}
\nonumber \lefteqn{\DKL(P_0(A_{1:s,i}), P_1(A_{1:s,i}) \mid a_{1:i-1}, b_{1:i-1}, s)} \\
\nonumber \phantom{+} &\stackrel{(i)}{\leq} \frac{1}{\tau({\emptyset})} \sum_{J} \Big(\frac{1-q}{q}\Big)^{2|J|} (S(J) - T(J))^2 \\
\nonumber  &\stackrel{(ii)}{\leq} (2-2q)^2 \Big( \Big(\frac{1-q}{q}\Big)^2 \sum_{j=1}^s (S({\{j\}}) - T({\{j\}}))^2 + 2 \sum_{|J| \geq 2} \Big(\frac{1-q}{q}\Big)^{2|J|} (S(J)^2 + T(J)^2) \\
\nonumber &\stackrel{(iii)}{=} \Big(\frac{m}{m-k+1}\Big)^s \Big(\Big(\frac{m}{m-2k+2}\Big)^2 \sum_{j=1}^s (S({\{j\}}) - T({\{j\}}))^2 + 2\sum_{|J| \geq 2} \Big(\frac{m}{m-2k+2}\Big)^{2|J|} (S(J)^2 + T(J)^2)\Big) \\
 &\stackrel{(iv)}{\leq} 3\sum_{j=1}^s (S(\{j\}) - T(\{j\}))^2 + 6\sum_{|J| \geq 2} \Big(\frac{m}{m-2k+2}\Big)^{2|J|} (S(J)^2 + T(J)^2).
\end{align}
Here (i) applies Proposition~\ref{prop:pb}, (ii) splits terms into the cases 
$|J| = 1$ and $|J| \geq 2$ (and applies the inequality $(S(J)-T(J))^2 \leq 2(S(J)^2 + T(J)^2)$ 
to the second set of terms), and (iii) substitutes in the definition of $q = \frac{1}{2} - \frac{k-1}{2(m-k+1)}$. 
Finally, (iv) uses the fact that $(\frac{m}{m-k+1})^s(\frac{m}{m-2k+2})^2 \leq \exp(s\frac{k-1}{m-k+1})\exp(2\frac{2k-2}{m-2k+2}) \leq \exp((2s+8)k/m)$ (as long as $k \leq m/4$). Then 
assuming $2s+8 \leq m/k$, this quantity is at most $\exp(1) \leq 3$ which yields the bound.

\section{Bounding Hypergeometric Expectations}
\label{sec:hg-details}

We show here that $\bE_{I \sim \HG(k-1,s,m)}[2^I-I-1] \leq \frac{4k^2s^2}{m^2}$, as 
required to bound \eqref{eq:I-pre}. This uses Theorem 4 of \citet{hoeffding1963probability}, 
which states that for any convex continuous function $f$, we have 
$\bE_{I \sim \HG(k-1,s,m)}[f(I)] \leq \bE_{I \sim \Binom(k-1, s/m)}[f(I)]$. 
The function $2^I - I - 1$ is certainly convex and continuous, so we have
\begin{align}
\bE_{I \sim \HG(k-1,s,m)}[2^I - I - 1]
 &\leq \bE_{I \sim \Binom(k-1,s/m)}[2^I - I- 1] \\
 &= \sum_{j=0}^{k-1} \binom{k-1}{j} (s/m)^j(1-s/m)^{k-1-j}(2^j-j-1) \\
 &\leq \sum_{j=2}^{k-1} \binom{k-1}{j} (s/m)^j 2^j \\
 &= (1 + 2s/m)^{k-1} - 2(k-1)s/m - 1 \leq 4(k-1)^2s^2/m^2 \leq 4k^2s^2/m^2
\end{align}
as long as $2(k-1)s/m \leq 1$.

\bibliographystyle{plainnat}
\bibliography{refdb/all}

\end{document}